\documentclass[aps,prb,longbibliography,superscriptaddress,preprint]{revtex4-1}
\usepackage{mathrsfs}
\usepackage{amsmath}
\usepackage{amsfonts}
\usepackage{amssymb}
\usepackage{amsthm}
\usepackage{graphicx}
\usepackage{color}
\usepackage{hyperref}
\usepackage{bm}
\usepackage[caption=false]{subfig}
\usepackage{verbatim}

\begin{document}
\title{Gilbert damping phenomenology for two-sublattice magnets}
\author{Akashdeep Kamra}
\email{akashdeep.kamra@ntnu.no} 
\affiliation{Center for Quantum Spintronics, Department of Physics, Norwegian University of Science and Technology, NO-7491 Trondheim, Norway}
\author{Roberto E. Troncoso}
\affiliation{Center for Quantum Spintronics, Department of Physics, Norwegian University of Science and Technology, NO-7491 Trondheim, Norway}
\author{Wolfgang Belzig}
\affiliation{Department of Physics, University of Konstanz, D-78457 Konstanz, Germany}
\author{Arne Brataas}
\email{arne.brataas@ntnu.no}
\affiliation{Center for Quantum Spintronics, Department of Physics, Norwegian University of Science and Technology, NO-7491 Trondheim, Norway}

\begin{abstract}
We present a systematic phenomenological description of Gilbert damping in two-sublattice magnets. Our theory covers the full range of materials from ferro- via ferri- to antiferromagnets. Following a Rayleigh dissipation functional approach within a  Lagrangian classical field formulation, the theory captures intra- as well as cross-sublattice terms in the Gilbert damping, parameterized by a 2$\times$2 matrix. When spin-pumping into an adjacent conductor causes dissipation, we obtain the corresponding Gilbert damping matrix in terms of the interfacial spin-mixing conductances. Our model reproduces the experimentally observed enhancement of the ferromagnetic resonance linewidth in a ferrimagnet close to its compensation temperature without requiring an increased Gilbert parameter. It also predicts new contributions to damping in an antiferromagnet and suggests the resonance linewidths as a direct probe of the sublattice asymmetry, which may stem from boundary or bulk. 
\end{abstract}


\maketitle

\section{Introduction}\label{sec:intro}
The fundamental connection~\cite{Barnett1935} between magnetic moment and spin angular momentum underlies the important role for magnets in nearly all spin-based concepts. An applied magnetic field provides the means to manipulate the state of a ferromagnet (FM), and thus the associated spin. Conversely, a spin-polarized current absorbed by the FM affects its magnetization~\cite{Slonczewski1996,Berger1996,Ralph2008,Brataas2012b}. Exploiting a related phenomenon, switching the state of an antiferromagnet (AFM) has also been achieved~\cite{Wadley2016}. Emboldened by this newly gained control, there has been an upsurge of interest in AFMs~\cite{Gomonay2014,Jungwirth2016,Baltz2018,Gomonay2018}, which offer several advantages over FMs. These include the absence of stray fields and a larger anisotropy-induced gap in the magnon spectrum. The two-sublattice nature of the AFMs further lends itself to phenomena distinct from FMs~\cite{Libor2018}. 

Concurrently, ferrimagnets (FiMs) have been manifesting their niche in a wide range of phenomena such as ultrafast switching~\cite{Hansteen2005,Stanciu2007,Graves2013} and low-dissipation spin transport~\cite{Uchida2010,Adachi2013,Kruglyak2010,Chumak2015,Bauer2012,Cornelissen2015,Goennenwein2015,Weiler2013}. A class of FiMs exhibits the so-called compensation temperature~\cite{Rodrigue1960,Geschwind1959,Gepraegs2016,Cramer2017,Niklas2017,Kim2018}, at which the net magnetization vanishes, similar to the case of AFMs. Despite a vanishing magnetization in the compensated state, most properties remain distinct from that of AFMs~\cite{Gurevich1996}. Thus, these materials can be tuned to mimic FMs and AFMs via the temperature. In conjunction with the possibility of a separate angular-momentum compensation, when the magnetization does not vanish but the total spin does, FiMs provide a remarkably rich platform for physics and applications. An increased complexity in the theoretical description~\cite{Gurevich1996,Kamra2017B} hence accompanies these structurally complicated materials, and may be held responsible for comparatively fewer theoretical studies. Nevertheless, a two-sublattice model with distinct parameters for each sublattice qualitatively captures all the phenomena mentioned above. 

Dissipation strongly influences the response of a magnet to a stimulus and is thus central to the study of magnetic phenomena such as switching, domain wall motion and spin transport. Nevertheless, magnetic damping has conventionally been investigated via the ferromagnetic resonance (FMR) linewidth. It is accounted for phenomenologically in the Landau-Lifshitz description of the magnetization dynamics via the so-called Gilbert damping term~\cite{Gilbert2004}, which produces a good agreement with experiments for a wide range of systems. The Gilbert damping represents the viscous contribution and may be `derived' within a Lagrangian formulation of classical field theory by including the Rayleigh dissipation functional~\cite{Gilbert2004}. While the magnetic damping for FMs has been studied in great detail~\cite{Gurevich1996,Akhiezer1968,Sparks1961,Gilbert2004,Tserkovnyak2002,Tserkovnyak2005}, from phenomenological descriptions to microscopic models, a systematic development of an analogous description for ferri- and antiferromagnets has been lacking in literature. Furthermore, recent theoretical results on spin pumping in two-sublattice magnets~\cite{Kamra2017} and damping in AFMs~\cite{Liu2017} suggest an important role for the previously disregarded~\cite{Gurevich1996} cross-sublattice terms in Gilbert damping, and thus set the stage for the present study. Yuan and co-workers have recently presented a step in this direction focussing on spin torques in AFMs~\cite{Yuan2018}.

Here, we formulate the magnetization dynamics equations in a general two-sublattice magnet following the classical Lagrangian approach that has previously been employed for FMs~\cite{Gilbert2004}. The Gilbert damping is included phenomenologically via a Rayleigh dissipation functional appropriately generalized to the two-sublattice system, which motivates intra- as well as cross-sublattice terms. The Gilbert damping parameter thus becomes a 2$\times$2 matrix, in contrast with its scalar form for a single-sublattice FM. Solving the system of equations for spatially homogeneous modes in a collinear ground state, we obtain the decay rates of the two eigenmodes finding direct pathways towards probing the dissipation mechanism and asymmetries in the system. Consistent with recent experiments~\cite{Hannes2017,Kim2018}, we find an enhancement in the decay rates~\cite{Hannes2017} close to the magnetization compensation in a FiM with an unaltered damping matrix~\cite{Kim2018}. The general description is found to be consistent with the spin pumping mediated damping in the magnet~\cite{Tserkovnyak2005,Kamra2017,Tserkovnyak2002}, and allows for relating the Gilbert damping matrix with the interfacial spin-mixing conductances. Focusing on AFMs, we express the magnetization dynamics in terms of the Neel variable thus clarifying the origin of the different damping terms in the corresponding dynamical equations~\cite{Hals2011,Yuan2018}. Apart from the usually considered terms, we find additional contributions for the case when sublattice-symmetry is broken in the AFM~\cite{Belashchenko2010,He2010,Kosub2017,Nogues1999,Kamra2017,Kamra2018}. Thus, FMR linewidth measurements offer a direct, parameter-free means of probing the sublattice asymmetry in AFMs, complementary to the spin pumping shot noise~\cite{Kamra2017}.

This paper is organized as follows. We derive the Landau-Lifshitz-Gilbert (LLG) equations for the two-sublattice model in Sec. \ref{sec:dynamics}. The ensuing equations are solved for the resonance frequencies and decay rates of the uniform modes in a collinear magnet in Sec. \ref{sec:uniform}. Section \ref{sec:special} presents the application of the phenomenology to describe a compensated ferrimagnet and spin pumping mediated Gilbert damping. The case of AFMs is discussed in Sec. \ref{sec:AFM}. We comment on the validity and possible generalizations of the theory in Sec. \ref{sec:discussion}. The paper is concluded with a summary in Sec. \ref{sec:summary}. The discussion of a generalized Rayleigh dissipation functional and properties of the damping matrix is deferred to the appendix.

\section{Magnetization dynamics and Gilbert damping}\label{sec:dynamics}
We consider a two-sublattice magnet described by classical magnetization fields $\pmb{M}_{A} \equiv \pmb{M}_{A}(\pmb{r},t)$ and $\pmb{M}_{B} \equiv \pmb{M}_{B}(\pmb{r},t)$ corresponding to the sublattices $A$ and $B$. The system is characterized by a magnetic free energy $F[\pmb{M}_A,\pmb{M}_B]$ with the magnetization fields assumed to be of constant magnitudes $M_{A0}$ and $M_{B0}$. Here, the notation $F[~]$ is employed to emphasize that the free energy is a functional over the magnetization fields, i.e. an integration of the free energy density over space. 

The undamped magnetization dynamics is described by equating the time derivative of the spin angular momentum associated with the magnetization to the torque experienced by it. The resulting Landau-Lifshitz equations for the two fields may be written as:
\begin{align}\label{eq:angflow}
\frac{d}{dt} \left( \frac{\pmb{M}_{A,B}}{- |\gamma_{A,B}|} \right)  =  - \frac{\dot{\pmb{M}}_{A,B}}{ |\gamma_{A,B}|}  = & ~\pmb{M}_{A,B} \times \mu_0 \pmb{H}_{A,B},
\end{align}
where $\gamma_{A,B} ~(< 0)$ are the gyromagnetic ratios for the two sublattices, and $\pmb{H}_{A,B}$ are the effective magnetic fields experienced by the respective magnetizations. This expression of angular momentum flow may be derived systematically within the Lagrangian classical field theory~\cite{Gilbert2004}. The same formalism also allows to account for a restricted form of damping via the so-called dissipation functional $R[\dot{\pmb{M}}_A,\dot{\pmb{M}}_B]$ in the generalized equations of motion:  
\begin{align}\label{eq:lang}
\frac{d}{dt} \frac{\delta \mathcal{L}[\cdot]}{\delta \dot{\pmb{M}}_{A,B}} - \frac{\delta \mathcal{L}[\cdot]}{\delta \pmb{M}_{A,B}} = & - \frac{\delta R[\dot{\pmb{M}}_A,\dot{\pmb{M}}_B]}{\delta \dot{\pmb{M}}_{A,B}},
\end{align}
where $\mathcal{L}[\cdot] \equiv \mathcal{L}[\pmb{M}_A,\pmb{M}_B,\dot{\pmb{M}}_A,\dot{\pmb{M}}_B] $ is the Lagrangian of the magnetic system. Here, $\delta \mathcal{L}[\cdot] / \delta \pmb{M}_{A}$ represents the functional derivative of the Lagrangian with respect to the various components of $\pmb{M}_A$, and so on. The left hand side of Eq. (\ref{eq:lang}) above represents the conservative dynamics of the magnet and reproduces Eq. (\ref{eq:angflow}) with~\cite{Gilbert2004}
\begin{align}\label{eq:Hab}
\mu_0 \pmb{H}_{A,B} = & - \frac{\delta F[\pmb{M}_A,\pmb{M}_B]}{\delta \pmb{M}_{A,B}},
\end{align}
while the right hand side accounts for the damping. 

The Gilbert damping is captured by a viscous Rayleigh dissipation functional parameterized by a symmetric matrix $\eta_{ij}$ with $\{ i,j\} = \{ A,B\}$:
\begin{align}
R[\dot{\pmb{M}}_{A},\dot{\pmb{M}}_{B}] = & \int_{V} d^3r \left( \frac{\eta_{AA}}{2} \dot{\pmb{M}}_{A} \cdot \dot{\pmb{M}}_{A} + \frac{\eta_{BB}}{2} \dot{\pmb{M}}_{B} \cdot \dot{\pmb{M}}_{B}  + \eta_{AB} \dot{\pmb{M}}_{A} \cdot \dot{\pmb{M}}_{B} \right),
\end{align}
where $V$ is the volume of the magnet. The above form of the functional assumes the damping to be spatially homogeneous, isotropic, and independent of the equilibrium configuration. A more general form with a lower symmetry is discussed in appendix \ref{sec:RD}. Including the dissipation functional via Eq. (\ref{eq:lang}) leads to the following replacements in the equations of motion (\ref{eq:angflow}):
\begin{align}
\mu_0 \pmb{H}_{A} \to \mu_0 \pmb{H}_{A} - \eta_{AA} \dot{\pmb{M}}_A - \eta_{AB} \dot{\pmb{M}}_B, \\
\mu_0 \pmb{H}_{B} \to \mu_0 \pmb{H}_{B} - \eta_{BB} \dot{\pmb{M}}_B - \eta_{AB} \dot{\pmb{M}}_A.
\end{align}
Hence, the LLG equations for the two-sublattice magnet become:
\begin{align}
\dot{\pmb{M}}_{A} = & - |\gamma_A| \left( \pmb{M}_A \times \mu_0 \pmb{H}_{A} \right) + |\gamma_A| \eta_{AA} \left( \pmb{M}_A \times \dot{\pmb{M}}_A \right) + |\gamma_A| \eta_{AB} \left( \pmb{M}_A \times \dot{\pmb{M}}_B \right), \\
\dot{\pmb{M}}_{B} = & - |\gamma_B| \left( \pmb{M}_B \times \mu_0 \pmb{H}_{B} \right) + |\gamma_B| \eta_{AB} \left( \pmb{M}_B \times \dot{\pmb{M}}_A \right) + |\gamma_B| \eta_{BB} \left( \pmb{M}_B \times \dot{\pmb{M}}_B \right).
\end{align}
These can further be expressed in terms of the unit vectors $\hat{\pmb{m}}_{A,B} = \pmb{M}_{A,B} / M_{A0,B0}$: 
\begin{align}
\dot{\hat{\pmb{m}}}_{A} = & - |\gamma_A| \left( \hat{\pmb{m}}_A \times \mu_0 \pmb{H}_{A} \right) + \alpha_{AA} \left( \pmb{m}_A \times \dot{\hat{\pmb{m}}}_A \right) + \alpha_{AB} \left( \hat{\pmb{m}}_A \times \dot{\hat{\pmb{m}}}_B \right), \label{eq:madot} \\
\dot{\hat{\pmb{m}}}_{B} = & - |\gamma_B| \left( \hat{\pmb{m}}_B \times \mu_0 \pmb{H}_{B} \right) + \alpha_{BA} \left( \hat{\pmb{m}}_B \times \dot{\hat{\pmb{m}}}_A \right) + \alpha_{BB} \left( \hat{\pmb{m}}_B \times \dot{\hat{\pmb{m}}}_B \right), \label{eq:mbdot}
\end{align}
thereby introducing the Gilbert damping matrix $\tilde{\alpha}$ for a two-sublattice system:
\begin{align}\label{eq:dampmat}
\tilde{\alpha} =   \begin{pmatrix}
\alpha_{AA} & \alpha_{AB} \\
\alpha_{BA} & \alpha_{BB}
\end{pmatrix}  = & \begin{pmatrix}
|\gamma_A| \eta_{AA} M_{A0} & |\gamma_A| \eta_{AB} M_{B0} \\
|\gamma_B| \eta_{AB} M_{A0} & |\gamma_B| \eta_{BB} M_{B0}
\end{pmatrix}, \\
\frac{\alpha_{AB}}{\alpha_{BA}} = & \frac{|\gamma_A| M_{B0}}{|\gamma_B| M_{A0}}. \label{eq:offratio}
\end{align}
As elaborated in appendix \ref{sec:DM}, the positivity of the dissipation functional implies that the eigenvalues and the determinant of $\tilde{\alpha}$ must be non-negative, which is equivalent to the following conditions:
\begin{align}\label{eq:det}
\eta_{AA}, \eta_{BB} \geq 0, \quad \eta_{AA} \eta_{BB} \geq \eta_{AB}^2 \implies \alpha_{AA}, \alpha_{BB} \geq 0, \quad \alpha_{AA} \alpha_{BB} \geq \alpha_{AB} \alpha_{BA} .
\end{align}
Thus, Eqs. (\ref{eq:madot}) and (\ref{eq:mbdot}) constitute the main result of this section, and introduce the damping matrix [Eq. (\ref{eq:dampmat})] along with the constraints imposed on it [Eq. (\ref{eq:offratio}) and (\ref{eq:det})] by the underlying formalism.

\section{Uniform modes in collinear ground state}\label{sec:uniform}
\begin{figure}[t]
	\begin{center}
		\includegraphics[height=100mm]{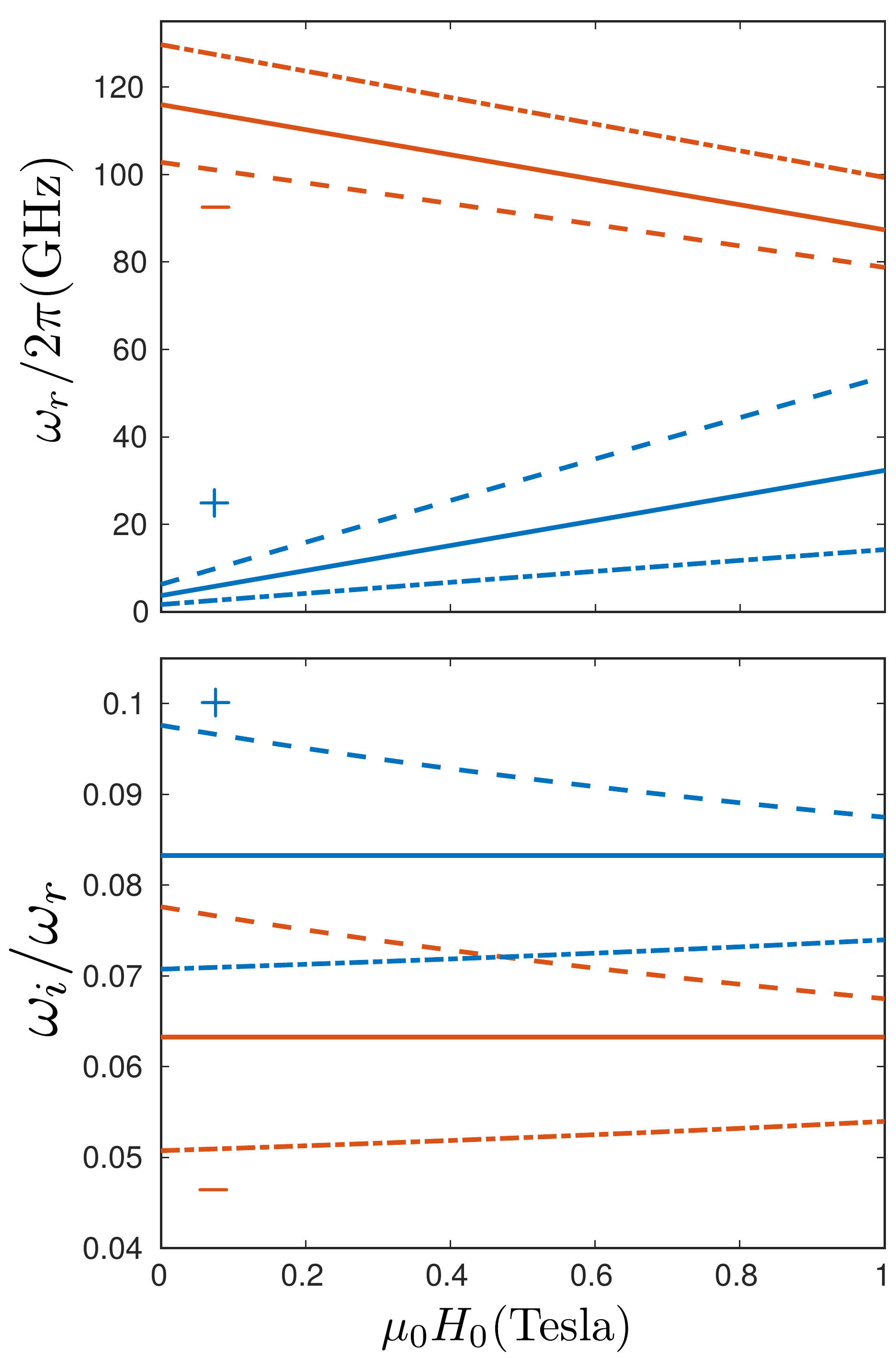}
		\caption{Resonance frequencies and normalized decay rates vs. the applied field for a quasi-ferromagnet ($M_{A0} = 5 M_{B0}$). $|\gamma_A|/|\gamma_B| = 1, 1.5, 0.5$ correspond to solid, dashed and dash-dotted lines respectively. The curves in blue and red respectively depict the $+$ and $-$ modes. The damping parameters employed are $\alpha_{AA} = 0.06$, $\alpha_{BB} = 0.04$ and $\alpha_{AB} = 0$.}
		\label{fig:field}
	\end{center}
\end{figure}

In this section, we employ the phenomenology introduced above to evaluate the resonance frequencies and the decay rates of the spatially homogeneous modes that can be probed in a typical FMR experiment. We thus work in the macrospin approximation, i.e. magnetizations are assumed to be spatially invariant. Considering an antiferromagnetic coupling $J~(> 0)$ between the two sublattices and parameterizing uniaxial easy-axis anisotropies via $K_{A,B} ~(>0)$, the free energy assumes the form:
\begin{align}\label{eq:free}
F[\pmb{M}_A,\pmb{M}_B] = &   \int_{V} d^3r  \left[ - \mu_0 H_0 (M_{Az} + M_{Bz}) - K_{A} M_{Az}^2 - K_{B} M_{Bz}^2  + J \pmb{M}_A \cdot \pmb{M}_{B} \right] ,
\end{align}
where $H_0 \hat{\pmb{z}}$ is the applied magnetic field. The magnet is assumed to be in a collinear ground state: $\pmb{M}_A = M_{A0} \hat{\pmb{z}}$ and $\pmb{M}_B = - M_{B0} \hat{\pmb{z}}$ with $M_{A0} > M_{B0}$. Employing Eq. (\ref{eq:Hab}) to evaluate the effective fields, the magnetization dynamics is expressed via the LLG equations (\ref{eq:madot}) and (\ref{eq:mbdot}). Considering $\pmb{M}_A =  M_{Ax} \hat{\pmb{x}} + M_{Ay} \hat{\pmb{y}} +  M_{A0} \hat{\pmb{z}}$, $\pmb{M}_B =  M_{Bx} \hat{\pmb{x}} + M_{By} \hat{\pmb{y}} -  M_{B0} \hat{\pmb{z}}$ with $|M_{Ax,Ay}| \ll M_{A0}$, $|M_{Bx,By}| \ll M_{B0}$, we linearize the resulting dynamical equations. Converting to Fourier space via $M_{Ax} = \mathcal{M}_{Ax} \exp{(i \omega t)}$ etc. and switching to circular basis via $\mathcal{M}_{A\pm (B\pm)} = \mathcal{M}_{Ax(Bx)} \pm i \mathcal{M}_{Ay(By)}$, we obtain two sets of coupled equations expressed succinctly as:
\begin{align}\label{eq:2x2}
\begin{pmatrix}
 \pm \omega - \Omega_A - i \omega \alpha_{AA} & - \left( |\gamma_A| J M_{A0} + i \omega \alpha_{AB} \frac{M_{A0}}{M_{B0}} \right)  \\
 \left( |\gamma_B| J M_{B0} + i \omega \alpha_{BA} \frac{M_{B0}}{M_{A0}} \right) &  \pm \omega + \Omega_B + i \omega \alpha_{BB}
\end{pmatrix} \begin{pmatrix}
\mathcal{M}_{A\pm} \\ \mathcal{M}_{B\pm}
\end{pmatrix} = & \begin{pmatrix}
0 \\ 0
\end{pmatrix},
\end{align}
where we define $\Omega_{A} \equiv |\gamma_A| (J M_{B0} + 2 K_A M_{A0} + \mu_0 H_0) $ and $\Omega_{B} \equiv |\gamma_B| (J M_{A0} + 2 K_B M_{B0} - \mu_0 H_0) $. Substituting $\omega = \omega_{r\pm} + i \omega_{i\pm}$ into the ensuing secular equation, we obtain the resonance frequencies $\omega_{r\pm}$ to the zeroth order and the corresponding decay rates $\omega_{i\pm}$ to the first order in the damping matrix elements:
\begin{align}
\omega_{r\pm} = & \frac{ \pm (\Omega_A - \Omega_B) + \sqrt{(\Omega_A + \Omega_B)^2 - 4 J^2 |\gamma_{A}| |\gamma_{B}| M_{A0} M_{B0}}}{2}, \label{eq:freqs} \\
\frac{\omega_{i \pm}}{\omega_{r\pm}} = &  \frac{\pm \omega_{r\pm} (\alpha_{AA} - \alpha_{BB}) + \alpha_{AA} \Omega_B + \alpha_{BB} \Omega_A - 2 J |\gamma_B| M_{A0} \alpha_{AB} }{\omega_{r+} + \omega_{r-}}  . \label{eq:decays}
\end{align}
In the expression above, Eq.\ (\ref{eq:freqs}) and Eq.\ (\ref{eq:decays}), we have chosen the positive solutions of the secular equations for the resonance frequencies. The negative solutions  are equal in magnitude to the positive ones and physically represent the same two modes. The positive-polarized mode in our notation corresponds to the typical ferromagnetic resonance mode, while the negative-polarized solution is sometimes termed `antiferromagnetic resonance'~\cite{Gepraegs2016}. In order to avoid confusion with the ferromagnetic or antiferromagnetic nature of the underlying material, we call the two resonances as positive- and negative-polarized. The decay rates can further be expressed in the following form:
\begin{align}\label{eq:decay2}
\frac{\omega_{i \pm}}{\omega_{r\pm}} = &  \frac{  \bar{\alpha}  \left(\Omega_A + \Omega_B \right) - 2 J |\gamma_B| M_{A0} \alpha_{AB} }{\omega_{r+} + \omega_{r-}} \pm \Delta \bar{\alpha} ,
\end{align}
with $\bar{\alpha} \equiv \left(\alpha_{AA} + \alpha_{BB}\right)/2$ and $\Delta \bar{\alpha} \equiv \left(\alpha_{AA} - \alpha_{BB}\right)/2$. Eq.\ (\ref{eq:decay2}) constitutes the main result of this section and demonstrates that (i) asymmetric damping in the two sublattices is manifested directly in the normalized decay rates of the two modes (Figs. \ref{fig:field} and \ref{fig:comp}), and (ii) off-diagonal components of the damping matrix may reduce the decay rates (Fig. \ref{fig:comp}). Furthermore, it is consistent with and reproduces the mode-dependence of the decay rates observed in the numerical studies of some metallic AFMs~\cite{Liu2017}.

To gain further insight into the results presented in Eqs. (\ref{eq:freqs}) and (\ref{eq:decay2}), we plot the resonance frequencies and the normalized decay rates vs. the applied magnetic field for a typical quasi-ferromagnet, such as yttrium iron garnet, in Fig. \ref{fig:field}. The parameters employed in the plot are $|\gamma_B| = 1.8 \times 10^{11}$, $M_{B0} = 10^5$, $K_{A} = K_{B} = 10^{-7}$, and $J = 10^{-5}$ in SI units, and have been chosen to represent the typical order of magnitude without pertaining to a specific material. The plus-polarized mode is lower in energy and is raised with an increasing applied magnetic field. The reverse is true for the minus-polarized mode whose relatively large frequency makes it inaccessible to typical ferromagnetic resonance experiments. As anticipated from Eq. (\ref{eq:decay2}), the normalized decay rates for the two modes differ when $\alpha_{AA} \neq \alpha_{BB}$. Furthermore, the normalized decay rates are independent of the applied field for symmetric gyromagnetic ratios for the two sublattices. Alternately, a measurement of the normalized decay rate for the plus-polarized mode is able to probe the sublattice asymmetry in the gyromagnetic ratios. Thus it provides essential information about the sublattices without requiring the measurement of the large frequency minus-polarized mode.

\section{Specific applications}\label{sec:special}
We now examine two cases of interest: (i) the mode decay rate in a ferrimagnet close to its compensation temperature, and (ii) the Gilbert damping matrix due to spin pumping into an adjacent conductor.   

\subsection{Compensated ferrimagnets}
FMR experiments carried out on gadolinium iron garnet~\cite{Rodrigue1960,Hannes2017} find an enhancement in the linewidth, and hence the mode decay rate, as the temperature approaches the compensation condition, i.e. when the two effective~\footnote{The garnets have a complicated unit cell with several magnetic ions. Nevertheless, the two-sublattice model employed here captures the essential physics.} sublattices have equal saturation magnetizations. These experiments have conventionally been interpreted in terms of an effective single-sublattice model thereby ascribing the enhancement in the decay rate to an increase in the scalar Gilbert damping constant allowed within the single-sublattice model~\cite{Geschwind1959}. In contrast, experiments probing the Gilbert parameter in a different FiM via domain wall velocity find it to be essentially unchanged around compensation~\cite{Kim2018}. Here, we analyze FMR in a compensated FiM using the two-sublattice phenomenology developed above and thus address this apparent inconsistency. 

\begin{figure}[t]
	\begin{center}
		\includegraphics[height=100mm]{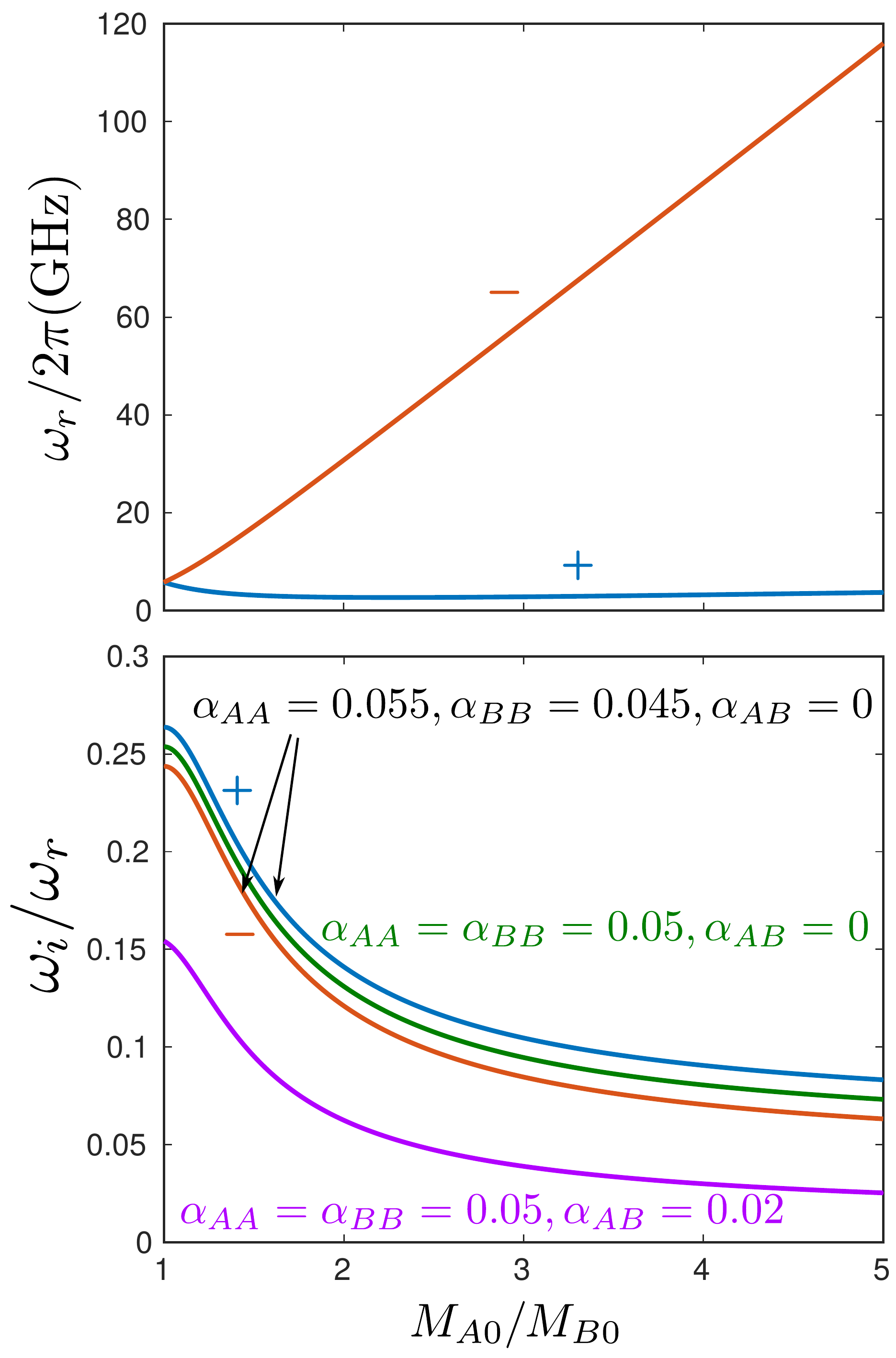}
		\caption{Resonance frequencies and normalized decay rates vs. relative saturation magnetizations of the sublattices. The curves which are not labeled as $+$ or $-$ represent the common normalized decay rates for both modes. The parameters employed are the same as for Fig. \ref{fig:field} with $\gamma_A = \gamma_B$.}
		\label{fig:comp}
	\end{center}
\end{figure}

The compensation behavior of a FiM may be captured within our model by allowing $M_{A0}$ to vary while keeping $M_{B0}$ fixed. The mode frequencies and normalized decay rates are examined with respect to the saturation magnetization variation in Fig. \ref{fig:comp}. We find an enhancement in the normalized decay rate, consistent with the FMR experiments~\cite{Rodrigue1960,Hannes2017}, for a fixed Gilbert damping matrix. The single-sublattice interpretation ascribes this change to a modification of the effective Gilbert damping parameter~\cite{Geschwind1959}, which is equal to the normalized decay rate within that model. In contrast, the latter is given by Eq. (\ref{eq:decay2}) within the two-sublattice model and evolves with the magnetization without requiring a modification in the Gilbert damping matrix. Specifically, the enhancement in decay rate observed at the compensation point is analogous to the so-called exchange enhancement of damping in AFMs~\cite{Keffer1952}. Close to compensation, the FiM mimics an AFM to some extent. 

We note that while the spherical samples employed in Ref. \onlinecite{Rodrigue1960} are captured well by our simple free energy expression [Eq. (\ref{eq:free})], the interfacial and shape anisotropies of the thin film sample employed in Ref. \onlinecite{Hannes2017} may result in additional contributions to decay rates. The similarity of the observed linewidth trends for the two kinds of samples suggests that these additional anisotropy effects may not underlie the observed damping enhancement. Quantitatively accounting for these thin film effects requires a numerical analysis, as discussed in Sec \ref{sec:discussion} below, and is beyond the scope of the present work. Furthermore, domain formation may result in additional damping contributions not captured within our single-domain model.

\subsection{Spin pumping mediated Gilbert damping}
Spin pumping~\cite{Tserkovnyak2002} from a FM into an adjacent conductor has been studied in great detail~\cite{Tserkovnyak2005} and has emerged as a key method for injecting pure spin currents into conductors~\cite{Maekawa2012}. The angular momentum thus lost into the conductor results in a contribution to the magnetic damping on top of the intrinsic dissipation in the bulk of the magnet. A variant of spin pumping has also been found to be the dominant cause of dissipation in metallic magnets~\cite{Liu2017}. Thus, we evaluate the Gilbert damping matrix arising due to spin pumping from a two-sublattice magnet~\cite{Kamra2017} into an adjacent conductor acting as an ideal spin sink. 

Within the macrospin approximation, the total spin contained by the magnet is given by:
\begin{align}
\pmb{S} = - \frac{M_{A0} V \hat{\pmb{m}}_A}{|\gamma_A|} - \frac{M_{B0} V \hat{\pmb{m}}_B}{|\gamma_B|}.
\end{align} 
The spin pumping current emitted by the two-sublattice magnet has the following general form~\cite{Kamra2017}:
\begin{align}
\pmb{I}_s = & \frac{\hbar}{e} \sum_{i,j = \{ A,B \}} G_{ij} \left( \hat{\pmb{m}}_i \times \dot{\hat{\pmb{m}}}_j \right),
\end{align}
with $G_{AB} = G_{BA}$, where the spin-mixing conductances $G_{ij}$ may be evaluated within different microscopic models~\cite{Cheng2014,Kamra2017,Sverre2018,Eirik2017}. Equating the spin pumping current to $- \dot{\pmb{S}}$ and employing Eqs. (\ref{eq:madot}) and (\ref{eq:mbdot}), the spin pumping contribution to the Gilbert damping matrix becomes:
\begin{align}
\alpha^\prime_{ij} & = \frac{\hbar G_{ij} |\gamma_i|}{e M_{i0} V},
\end{align} 
which in turn implies
\begin{align}
\eta^\prime_{ij} = & \frac{\hbar G_{ij}}{e M_{i0} M_{j0} V},
\end{align}
for the corresponding dissipation functional. The resulting Gilbert damping matrix is found to be consistent with its general form and constraints formulated in Sec. \ref{sec:dynamics}. Thus, employing the phenomenology developed above, we are able to directly relate the magnetic damping in a two-sublattice magnet to the spin-mixing conductance of its interface with a conductor.

\section{Antiferromagnets}\label{sec:AFM}
Due to their special place with high symmetry in the two-sublattice model as well as the recent upsurge of interest~\cite{Gomonay2014,Jungwirth2016,Baltz2018,Gomonay2018,Johansen2018,Alireza2018,Alireza2018B}, we devote the present section to a focused discussion on AFMs in the context of the general results obtained above. It is often convenient to describe the AFM in terms of a different set of variables:
\begin{align}\label{eq:mn}
\pmb{m} = \frac{\hat{\pmb{m}}_A + \hat{\pmb{m}}_B}{2}, \quad \pmb{n} = \frac{\hat{\pmb{m}}_A - \hat{\pmb{m}}_B}{2}.
\end{align} 
In contrast with $\hat{\pmb{m}}_A$ and $\hat{\pmb{m}}_B$, $\pmb{m}$ and $\pmb{n}$ are not unit vectors in general. The dynamical equations for $\pmb{m}$ and $\pmb{n}$ may be formulated by developing the entire field theory, starting with the free energy functional, in terms of $\pmb{m}$ and $\pmb{n}$. Such a formulation, including damping, has been accomplished by Hals and coworkers~\cite{Hals2011}. Here, we circumvent such a repetition and directly express the corresponding dynamical equations by employing Eqs. (\ref{eq:madot}) and (\ref{eq:mbdot}) into Eq. (\ref{eq:mn}):
\begin{align}
\dot{\pmb{m}} = & - \left(\pmb{m} \times \gamma_m \mu_0 \pmb{H}_m \right) - \left(\pmb{n} \times \gamma_n \mu_0 \pmb{H}_n \right) + \sum_{p,q = \{ m,n \} } \alpha^m_{pq} \left( \pmb{p} \times \dot{\pmb{q}} \right), \\
\dot{\pmb{n}} = & - \left(\pmb{m} \times \gamma_n \mu_0 \pmb{H}_n \right) - \left(\pmb{n} \times \gamma_m \mu_0 \pmb{H}_m \right) + \sum_{p,q = \{m,n\}} \alpha^n_{pq} \left( \pmb{p} \times \dot{\pmb{q}} \right), 
\end{align}
with
\begin{align}
\gamma_m \mu_0 \pmb{H}_m \equiv & \frac{|\gamma_A| \mu_0 \pmb{H}_A + |\gamma_B| \mu_0 \pmb{H}_B}{2}, \\
\gamma_n \mu_0 \pmb{H}_n \equiv & \frac{|\gamma_A| \mu_0 \pmb{H}_A - |\gamma_B| \mu_0 \pmb{H}_B}{2}, \\
\alpha^{m}_{mm} = \alpha^{n}_{nm} = & \frac{\alpha_{AA} + \alpha_{BB} + \alpha_{AB} + \alpha_{BA}}{2}, \\
\alpha^{m}_{mn} = \alpha^{n}_{nn} = & \frac{\alpha_{AA} - \alpha_{BB} - \alpha_{AB} + \alpha_{BA}}{2}, \\
\alpha^{m}_{nn} = \alpha^{n}_{mn} = & \frac{\alpha_{AA} + \alpha_{BB} - \alpha_{AB} - \alpha_{BA}}{2}, \\
\alpha^{m}_{nm} = \alpha^{n}_{mm} = & \frac{\alpha_{AA} - \alpha_{BB} + \alpha_{AB} - \alpha_{BA}}{2}.
\end{align}
A general physical significance, analogous to $\gamma_{A,B}$, may not be associated with $\gamma_{m,n}$ which merely serve the purpose of notation here. The equations obtained above manifest new damping terms in addition to the ones that are typically considered in the description of AFMs. Accounting for the sublattice symmetry of the antiferromagnetic bulk while allowing for the damping to be asymmetric, we may assume $\gamma_A = \gamma_B$ and $M_{A0} = M_{B0}$, with $\bar{\alpha} \equiv \left(\alpha_{AA} + \alpha_{BB}\right)/2$, $\Delta \bar{\alpha} \equiv \left(\alpha_{AA} - \alpha_{BB}\right)/2$, and $\alpha_{AB} = \alpha_{BA} \equiv \alpha_{od}$. Thus, the damping parameters simplify to
\begin{align}
\alpha^{m}_{mm} = \alpha^{n}_{nm} = & \bar{\alpha} + \alpha_{od}, \\
\alpha^{m}_{mn} = \alpha^{n}_{nn} = & \Delta \bar{\alpha}, \\
\alpha^{m}_{nn} = \alpha^{n}_{mn} = & \bar{\alpha} - \alpha_{od}, \\
\alpha^{m}_{nm} = \alpha^{n}_{mm} = & \Delta \bar{\alpha},
\end{align}
thereby eliminating the ``new'' terms in the damping when $\alpha_{AA} = \alpha_{BB}$. However, the sublattice symmetry may not be applicable to AFMs, such as FeMn, with non-identical sublattices. Furthermore, the sublattice symmetry of the AFM may be broken at the interface~\cite{Belashchenko2010,He2010,Kosub2017} via, for example, spin mixing conductances~\cite{Kamra2017,Kamra2018,Bender2017} resulting in $\alpha_{AA} \neq \alpha_{BB}$.

The resonance frequencies and normalized decay rates [Eqs. (\ref{eq:freqs}) and (\ref{eq:decay2})] take a simpler form for AFMs. Substituting $K_A = K_B \equiv K$, $\gamma_A = \gamma_B \equiv \gamma$, and $M_{A0} = M_{B0} \equiv M_0$:  
\begin{align}
\omega_{r\pm} = &  \pm |\gamma| \mu_0 H_0 + 2 |\gamma| M_0 \sqrt{ (J + K) K }, \\
\frac{\omega_{i \pm}}{\omega_{r\pm}} = &  \frac{ J (\bar{\alpha} - \alpha_{od})  + 2 K \bar{\alpha} }{2 \sqrt{ (J + K) K } } \pm \Delta \bar{\alpha} \approx \frac{(\bar{\alpha} - \alpha_{od})}{2} \sqrt{\frac{J}{K}} + \bar{\alpha} \sqrt{\frac{K}{J}} \pm \Delta \bar{\alpha},
\end{align}
where we have employed $J \gg K$ in the final simplification. The term $\propto \sqrt{K/J}$ has typically been disregarded on the grounds $K \ll J$. However, recent numerical studies of damping in several AFMs~\cite{Liu2017} find $\bar{\alpha} \gg \bar{\alpha} - \alpha_{od} > 0$ thus suggesting that this term should be comparable to the one proportional to $\sqrt{J/K}$ and hence may not be disregarded. The expression above also suggests measurement of the normalized decay rates as a means of detecting the sublattice asymmetry in damping. For AFMs symmetrical in the bulk, such an asymmetry may arise due to the corresponding asymmetry in the interfacial spin-mixing conductance~\cite{Kamra2017,Kamra2018,Bender2017}. Thus, decay rate measurements offer a method to detect and quantify such interfacial effects complementary to the spin pumping shot noise measurements suggested earlier~\cite{Kamra2017}.

\section{Discussion}\label{sec:discussion}
We have presented a phenomenological description of Gilbert damping in two-sublattice magnets and demonstrated how it can be exploited to describe and characterize the system effectively. We now comment on the limitations and possible generalizations of the formalism presented herein. To begin with, the two-sublattice model is the simplest description of ferri- and antiferromagnets. It has been successful in capturing a wide range of phenomenon. However, recent measurements of magnetization dynamics in nickel oxide could only be explained using an eight-sublattice model~\cite{Zhe2018}. The temperature dependence of the spin Seebeck effect in yttrium iron garnet also required accounting for more than two magnon bands~\cite{Barker2016}. A generalization of our formalism to a N-sublattice model is straightforward  and can be achieved via a Rayleigh dissipation functional with $\mathrm{N}^2$ terms, counting $\eta_{ij}$ and $\eta_{ji}$ as separate terms. The ensuing Gilbert damping matrix will be N$\times$N while obeying the positive determinant constraint analogous to Eq. (\ref{eq:det}).

In our description of the collinear magnet [Eq. (\ref{eq:free})], we have disregarded contributions to the free energy which break the uniaxial symmetry of the system about the z-axis. Such terms arise due to spin-nonconserving interactions~\cite{Kamra2016A}, such as dipolar fields and magnetocrystalline anisotropies, and lead to a mixing between the plus- and minus-polarized modes~\cite{Kamra2017B}. Including these contributions converts the two uncoupled 2$\times$2 matrix equations [(\ref{eq:2x2})] into a single 4$\times$4 matrix equation rendering the solution analytically intractable. A detailed analysis of these contributions~\cite{Kamra2017B} shows that their effect is most prominent when the two modes are quasi-degenerate, and may be disregarded in a first approximation. 

In evaluating the resonance frequencies and the decay rates [Eqs. (\ref{eq:freqs}) and (\ref{eq:decay2})], we have assumed the elements of the damping matrix to be small. A precise statement of the assumption employed is $\omega_{i} \ll \omega_{r}$, which simply translates to $\alpha \ll 1$ for a single-sublattice ferromagnet. In contrast, the constraint imposed on the damping matrix within the two-sublattice model by the assumption of small normalized decay rate is more stringent [Eq. (\ref{eq:decay2})]. For example, this assumption for an AFM with $\alpha_{AB} = \Delta \bar{\alpha} = 0$ requires $\bar{\alpha} \ll \sqrt{K/J} \ll 1$. This stringent constraint may not be satisfied in most AFMs~\cite{Liu2017}, thereby bringing the simple Lorentzian shape description of the FMR into question. It can also be seen from Fig. \ref{fig:comp} that the assumption of a small normalized decay rate is not very good for the chosen parameters.

\section{Summary}\label{sec:summary}
We have developed a systematic phenomenological description of the Gilbert damping in a two-sublattice magnet via inclusion of a Rayleigh dissipation functional within the Lagrangian formulation of the magnetization dynamics. Employing general expressions based on symmetry, we find cross-sublattice Gilbert damping terms in the LLG equations in consistence with other recent findings~\cite{Kamra2017,Liu2017,Yuan2018}. Exploiting the phenomenology, we explain the enhancement of damping~\cite{Rodrigue1960,Hannes2017} in a compensated ferrimagnet without requiring an increase in the damping parameters~\cite{Kim2018}. We also demonstrate approaches to probe the various forms of sublattice asymmetries. Our work provides a unified description of ferro- via ferri- to antiferromagnets and allows for understanding a broad range of materials and experiments that have emerged into focus in the recent years.

\section*{Acknowledgments}
A. K. thanks Hannes Maier-Flaig and Kathrin Ganzhorn for valuable discussions. We acknowledge financial support from the Alexander von Humboldt Foundation, the Research Council of Norway through its Centers of Excellence funding scheme, project 262633, ``QuSpin'', and the DFG through SFB 767 and SPP 1538.

\appendix

\section{Generalized Rayleigh dissipation functional}\label{sec:RD}
As compared to the considerations in Sec. \ref{sec:dynamics}, a more general approach to parameterizing the dissipation functional is given by:
	\begin{align}
	R[\dot{\pmb{M}}_{A},\dot{\pmb{M}}_{B}] = & \frac{1}{2} \int_{V} \int_{V} d^3r^\prime d^3r \sum_{p,q = \{ A,B \}} \sum_{i,j = \{ x,y,z \} } \dot{M}_{pi}(\pmb{r}) \eta_{pq}^{ij}(\pmb{r},\pmb{r}^\prime)  \dot{M}_{qj} (\pmb{r}^\prime) .
	\end{align}
This form allows to capture the damping in an environment with a reduced symmetry. However, the larger number of parameters also makes it difficult to extract them reliably via typical experiments. The above general form reduces to the case considered in Sec. \ref{sec:dynamics} when $\eta_{pq}^{ij}(\pmb{r},\pmb{r}^\prime) = \eta_{pq} \delta_{ij} \delta(\pmb{r} - \pmb{r}^\prime)$ and $\eta_{pq} = \eta_{qp}$. Furthermore, the coefficients $\eta_{pq}^{ij}$ may depend upon $\pmb{M}_{A}(\pmb{r})$ and $\pmb{M}_{B}(\pmb{r})$ as has been found in recent numerical studies of Gilbert damping in AFMs~\cite{Liu2017}.

\section{Damping matrix}\label{sec:DM}
The Rayleigh dissipation functional considered in the main text is given by:
\begin{align}
R[\dot{\pmb{M}}_{A},\dot{\pmb{M}}_{B}] = & \int_{V} d^3r \left( \frac{\eta_{AA}}{2} \dot{\pmb{M}}_{A} \cdot \dot{\pmb{M}}_{A} + \frac{\eta_{BB}}{2} \dot{\pmb{M}}_{B} \cdot \dot{\pmb{M}}_{B}  + \eta_{AB} \dot{\pmb{M}}_{A} \cdot \dot{\pmb{M}}_{B} \right),
\end{align}
which may be brought into the following concise form with the notation $\tilde{\dot{\pmb{M}}} \equiv [\dot{\pmb{M}}_A \ \dot{\pmb{M}}_B ]^\intercal$:
\begin{align}
R[\dot{\pmb{M}}_{A},\dot{\pmb{M}}_{B}] = & \frac{1}{2} \int_{V} d^3r \ \tilde{\dot{\pmb{M}}}^\intercal \ \tilde{\eta} \ \tilde{\dot{\pmb{M}}}, 
\end{align}
where $\tilde{\eta}$ is the appropriate matrix given by:
\begin{align}
\tilde{\eta} = &  \begin{pmatrix}
\eta_{AA} & \eta_{AB} \\
\eta_{AB} & \eta_{BB}
\end{pmatrix}.
\end{align}
Considering an orthogonal transformation $\tilde{\dot{\pmb{M}}} = \tilde{Q} \tilde{\dot{\mathcal{M}}}$, the dissipation functional can be brought to a diagonal form
\begin{align}
R[\dot{\pmb{M}}_{A},\dot{\pmb{M}}_{B}] = & \frac{1}{2} \int_{V} d^3r \ \tilde{\dot{\mathcal{M}}}^\intercal \ \tilde{Q}^\intercal \tilde{\eta} \tilde{Q} \ \tilde{\dot{\mathcal{M}}}, 
\end{align}
where $\tilde{Q}^\intercal \tilde{\eta} \tilde{Q}$ is assumed to be diagonal. The positivity of the dissipation for arbitrary magnetization dynamics then requires the two diagonal elements to be non-negative which further entails the non-negativity of the determinant of $\tilde{\eta}$:
\begin{align}
| \tilde{Q}^\intercal \tilde{\eta} \tilde{Q} | & \geq 0 , \\
| \tilde{Q}^\intercal| |\tilde{\eta}| |\tilde{Q} | & \geq 0, \\
|\tilde{\eta}| & \geq 0 \quad \implies \eta_{AA} \eta_{BB} \geq \eta_{AB}^2.
\end{align}

\bibliography{Gilbert_Ferri.bib}

\end{document}